\documentclass{amsart}
\usepackage{amssymb}
\usepackage{amsfonts}

\theoremstyle{plain}
\newtheorem{acknowledgement}{Acknowledgement}

\numberwithin{equation}{section}
\input{tcilatex}

\begin{document}
\title[Weighing the Milky Way]{Weighing the Milky Way}
\author{Munawar Karim}
\address{Department of Physics, St. John Fisher College, Rochester, NY 14618}
\email{karim@sjfc.edu}
\author{Angelo Tartaglia}
\address{Department of Physics, Politechnico, Turin, Italy}
\email{tartaglia@polito.it}
\author{Ashfaque H. Bokhari}
\address{Department of Physics, St. John Fisher College, Rochester, NY 14618}

\begin{abstract}
We describe an experiment to measure the mass of the Milky Way galaxy. The
experiment is based on calculated light travel times along orthogonal
directions in the Schwarzschild metric of the Galactic center. We show that
the difference is proportional to the Galactic mass. We apply the result to
light travel times in a 10cm Michelson type interferometer located on Earth.
The mass of the Galactic center is shown to contribute $10^{-6}$ to the flat
space component of the metric. An experiment is proposed to measure the
effect.

\begin{acknowledgement}
We acknowledge the technical advice of M. Bocko, and A. Marnell for help
with graphics. A.B. thanks St. John Fisher College for its hospitality and
the United States Fulbright Scholarship Board for the award of a Fulbright
Fellowship.
\end{acknowledgement}
\end{abstract}

\maketitle

\section{Introduction}

What is the mass of the Galaxy? We answer this question by computing, in a
model metric of the Galaxy, the travel time of light beams in a Michelson
type interferometer located on Earth's equator. We show that the light
travel time is proportional to the mass of \ the Galaxy. The result suggests
that an experiment can measure the mass of the Galaxy.

We start with a metric in Boyer-Lindquist coordinates, representing on the
equatorial plane, the axially symmetric field of a rotating matter
distribution with the same symmetry

\begin{equation}
ds^{2}=g_{tt}dt^{2}+2g_{t\phi }dtd\phi +g_{rr}dr^{2}+g_{\theta \theta
}d\theta ^{2}+g_{\phi \phi }d\phi ^{2}
\end{equation}

We will calculate and compare the light travel times for beams traveling
along the radial, polar and azimuthal axes. The interferometer is small
compared to the scale factors of the field.

\subsection{Light beam traveling along the $\protect\phi -$direction}

In this section we calculate the travel time of a beam of light traveling
along the $\phi -$direction. The beam is located on Earth's equator $\left(
\theta =\pi /2\right) $, a coordinate distance $R$ from the field center.
The interferometer orbits about the field center with a constant coordinate
speed $v=R\Omega $, where $\Omega $ is the coordinate angular speed. The
trajectory of light is a null geodesic, furthermore, $dr=d\theta =0$:

\begin{equation}
ds^{2}=0=g_{tt}dt^{2}+2g_{t\phi }dtd\phi +g_{\phi \phi }d\phi ^{2}
\end{equation}

We neglect the difference between a null geodesic and an arc of an
equatorial circle (valid when the coordinate arm length of the
interferometer $l$ $\ll $ $R$). The world line of the light beam between the
two mirrors is:

\begin{equation}
t_{\pm }=\frac{-g_{t\phi }\pm \sqrt{g_{t\phi }^{2}-g_{tt}g_{\phi \phi }}}{%
g_{tt}}\left( \phi -\phi _{0}\right)
\end{equation}

We choose $\phi _{0}=0$. The positive sign is used when the light beam is
traveling in the same sense as the interferometer. The world line of the end
mirror of the equatorial arm is:

\begin{equation}
t=\frac{\phi -\Phi }{\Omega }
\end{equation}%
where $\Phi $ is the angle subtended by the interferometer at the field
center. The light ray reaches the mirror when

\begin{equation}
\frac{-g_{t\phi }+\sqrt{g_{t\phi }^{2}-g_{tt}g_{\phi \phi }}}{g_{tt}}\phi =%
\frac{\phi -\Phi }{\Omega }
\end{equation}%
Solving for the azimuth gives

\begin{equation}
\phi _{r}=\frac{\Phi g_{tt}}{g_{t\phi }\Omega +g_{tt}-\Omega \sqrt{g_{t\phi
}^{2}-g_{tt}g_{\phi \phi }}}
\end{equation}%
and the coordinate time

\begin{equation}
t_{r}=-\Phi \frac{g_{t\phi }-\sqrt{g_{t\phi }^{2}-g_{tt}g_{\phi \phi }}}{%
g_{t\phi }\Omega +g_{tt}-\Omega \sqrt{g_{t\phi }^{2}-g_{tt}g_{\phi \phi }}}
\end{equation}%
The world line of the reflected ray is

\begin{equation}
t=t_{r}+\frac{-g_{t\phi }-\sqrt{g_{t\phi }^{2}-g_{tt}g_{\phi \phi }}}{g_{tt}}%
\left( \phi -\phi _{r}\right)
\end{equation}%
and of the beam splitter is

\begin{equation}
t=\frac{\phi }{\Omega }
\end{equation}%
The ray returns to the beam splitter when

\begin{equation}
\frac{\phi }{\Omega }=t_{r}+\frac{-g_{t\phi }-\sqrt{g_{t\phi
}^{2}-g_{tt}g_{\phi \phi }}}{g_{tt}}\left( \phi -\phi _{r}\right)
\end{equation}%
or when the azimuth is

\begin{equation}
\phi _{l}=\Omega \frac{g_{tt}t_{r}+g_{t\phi }\phi _{r}+\sqrt{g_{t\phi
}^{2}-g_{tt}g_{\phi \phi }}\phi _{r}}{g_{tt}+g_{t\phi }\Omega +\Omega \sqrt{%
g_{t\phi }^{2}-g_{tt}g_{\phi \phi }}}
\end{equation}%
and coordinate time%
\begin{equation}
t_{l}=\frac{g_{tt}t_{r}+g_{t\phi }\phi _{r}+\sqrt{g_{t\phi
}^{2}-g_{tt}g_{\phi \phi }}\phi _{r}}{g_{tt}+g_{t\phi }\Omega +\Omega \sqrt{%
g_{t\phi }^{2}-g_{tt}g_{\phi \phi }}}
\end{equation}%
Substituting the expressions for $\phi _{r}$ and $t_{r}$ gives

\begin{equation}
t_{l}=2\Phi \frac{\sqrt{g_{t\phi }^{2}-g_{tt}g_{\phi \phi }}}{%
g_{tt}+2g_{t\phi }\Omega +\Omega ^{2}g_{\phi \phi }}
\end{equation}%
In terms of proper time of the interferometer this is equivalent to

\begin{eqnarray}
\tau _{e} &=&\frac{1}{c}\sqrt{g_{tt}+2g_{t\phi }\Omega +\Omega ^{2}g_{\phi
\phi }}t_{l} \\
\tau _{e} &=&2\frac{\Phi }{c}\sqrt{\frac{g_{t\phi }^{2}-g_{tt}g_{\phi \phi }%
}{g_{tt}+2g_{t\phi }\Omega +\Omega ^{2}g_{\phi \phi }}}
\end{eqnarray}

\subsection{Light travel time along the $r-$direction}

We repeat the calculation for a light beam traveling inward from the
beam-splitter at $r=R$ to the end mirror a coordinate distance $l$ at $%
r=R^{\prime }$, where $R>R^{\prime }$. In this case $d\theta =0$. The
coordinate length of the interferometer arm is $l$. The world line of the
mirror is

\begin{eqnarray}
t &=&\frac{\phi }{\Omega } \\
r &=&R^{\prime }
\end{eqnarray}%
The world line of the light ray is now

\begin{equation}
t=\dint\limits_{R}^{R^{\prime }}\left\{ \frac{-g_{t\phi }\left( \frac{d\phi 
}{dr}\right) -\sqrt{g_{t\phi }^{2}\left( \frac{d\phi }{dr}\right)
^{2}-g_{tt}\left( g_{rr}+g_{\phi \phi }\left( \frac{d\phi }{dr}\right)
^{2}\right) }}{g_{tt}}\right\} dr
\end{equation}%
The negative sign of the radical is used because the light beam is traveling
inwards. Assume for the moment that $\frac{d\phi }{dr}$ is small enough to
be kept to first order only. The expression simplifies to:

\begin{equation}
t\simeq \dint\limits_{R}^{R^{\prime }}\left( -\frac{g_{t\phi }}{g_{tt}}\frac{%
d\phi }{dr}-\sqrt{-\frac{g_{rr}}{g_{tt}}}\right) dr
\end{equation}%
In the example under consideration we can approximate the space trajectory
of the light beam as a straight line starting from the origin at $\left(
R,0,0\right) $ and passing through $\left( R^{\prime },0,\phi _{i}\right) $.
We can relate $r$ with $\phi $

\begin{equation}
r=\frac{R}{\cos \phi +\tan \xi \sin \phi }
\end{equation}%
$\xi $ is fixed by imposing

\begin{equation}
R^{\prime }=\frac{R}{\cos \phi _{i}+\tan \xi \sin \phi _{i}}
\end{equation}%
This gives%
\begin{eqnarray}
\tan \xi &=&\frac{R-R^{\prime }\cos \phi _{i}}{R^{\prime }\sin \phi _{i}}%
\approx \frac{R-R^{\prime }}{R^{\prime }\phi _{i}}\text{ for }\phi _{i}\ll 1
\\
&\therefore &r=\frac{R}{\cos \phi +\frac{R-R^{\prime }\cos \phi _{i}}{%
R^{\prime }\sin \phi _{i}}\sin \phi } \\
r &\approx &R-R\phi \tan \xi +\left( \frac{1}{2}R\tan ^{2}\xi \right) \phi
^{2}+O\left( \phi ^{3}\right) \approx R-R\phi \tan \xi \\
\frac{d\phi }{dr} &\simeq &-\frac{1}{R\tan \xi }\simeq -\frac{1}{R}\frac{%
R^{\prime }}{R-R^{\prime }}\phi _{i}
\end{eqnarray}%
The expression for $t$ approximates to:

\begin{equation}
t\simeq \frac{1}{R}\frac{R^{\prime }}{R-R^{\prime }}\phi
_{i}\dint\limits_{R}^{R^{\prime }}\frac{g_{t\phi }}{g_{tt}}%
dr-\dint\limits_{R}^{R^{\prime }}\sqrt{-\frac{g_{rr}}{g_{tt}}}dr
\end{equation}

The metric elements are

\begin{eqnarray}
g_{tt} &=&c^{2}\left( 1-\frac{\alpha }{r}\right) \\
g_{rr} &=&-\left( 1-\frac{\alpha }{r}\right) ^{-1} \\
g_{\theta \theta } &=&-r^{2} \\
g_{\phi \phi } &=&-r^{2}\sin ^{2}\theta \\
g_{t\phi } &=&c\frac{\alpha }{r}a\sin ^{2}\theta
\end{eqnarray}

where $\alpha $ is the Schwarzschild radius and $a$ is the specific angular
momentum of the field source.

The integral for $t$ is:

\begin{eqnarray}
t &\simeq &\frac{1}{R}\frac{R^{\prime }}{R-R^{\prime }}\phi _{i}\frac{\alpha
a}{c}\dint\limits_{R}^{R^{\prime }}\frac{1}{r}dr-\frac{1}{c}%
\dint\limits_{R}^{R^{\prime }}\left( 1+\frac{\alpha }{r}\right) dr \\
t_{i} &\simeq &\frac{R-R^{\prime }}{c}+\frac{\alpha }{c}\left( 1-\frac{%
R^{\prime }\phi _{i}}{R\left( R-R^{\prime }\right) }a\right) \log \frac{R}{%
R^{\prime }}
\end{eqnarray}

Recalling that $\phi _{i}=\Omega t_{i}$

\begin{eqnarray}
t_{i} &\simeq &\frac{R-R^{\prime }}{c}+\frac{\alpha }{c}\left( 1-\frac{%
R^{\prime }\Omega t_{i}}{R\left( R-R^{\prime }\right) }a\right) \log \frac{R%
}{R^{\prime }} \\
t_{i} &\simeq &\frac{R-R^{\prime }}{c}+\frac{\alpha }{c}\left( 1-\frac{%
R^{\prime }\Omega t_{i}}{R}\frac{a}{c}\right) \log \frac{R}{R^{\prime }}
\end{eqnarray}

Under conditions where the length of the interferometer arm $l\ll R$

\begin{equation}
t_{i}\simeq \frac{l}{c}\left[ 1+\frac{\alpha }{R}\left( 1-\Omega \frac{a}{c}%
\right) \right]
\end{equation}

This expression reproduces the Shapiro time delay \cite{shapiro} with a
small correction due to the angular momentum of the field source.

\qquad For the return trip to the center of the interferometer,

\begin{equation}
r=\frac{R^{\prime }}{\cos \phi -\tan \zeta \sin \phi }
\end{equation}

In this case both $\phi $ and $t$ are zero at the starting point i.e., the
end mirror. Once again%
\begin{eqnarray}
\tan \zeta &=&\frac{R\cos \phi _{e}-R^{\prime }}{R\sin \phi _{e}}\simeq 
\frac{R-R^{\prime }}{R\phi _{e}} \\
\frac{d\phi }{dr} &\simeq &\frac{1}{R^{\prime }\tan \zeta }
\end{eqnarray}

The time for the return trip (using now a positive sign for the radical in
(1.18)), $t_{e}\simeq t_{i}$. The round-trip takes

\begin{equation}
t_{r}=t_{i}+t_{e}\simeq 2\frac{l}{c}\left[ 1+\frac{\alpha }{R}\left(
1-\Omega \frac{a}{c}\right) \right]
\end{equation}

We note that the corrections add in the round-trip travel time.

In terms of proper time

\begin{eqnarray}
\tau _{r} &=&\frac{1}{c}\sqrt{g_{tt}+2g_{t\phi }\Omega +g_{\phi \phi }\Omega
^{2}}t_{r} \\
\tau _{r} &\simeq &\sqrt{1-\frac{\alpha }{R}+2\frac{\alpha a}{R^{2}}\frac{%
R\Omega }{c}-\left( \frac{R\Omega }{c}\right) ^{2}}2\frac{l}{c}\left[ 1+%
\frac{\alpha }{R}\left( 1-\Omega \frac{a}{c}\right) \right]
\end{eqnarray}

We define three dimensionless parameters:

\begin{eqnarray}
\mu &\equiv &\frac{\alpha }{R} \\
\kappa &\equiv &\frac{a}{R} \\
\beta &\equiv &\frac{\Omega R}{c}
\end{eqnarray}

Keeping terms which are of lowest order in the parameters $\mu $, $\kappa $
and $\beta $, the radial proper time of flight is

\begin{equation}
\tau _{r}\simeq 2\frac{l}{c}\left[ 1+\frac{1}{2}\mu -\frac{1}{2}\beta ^{2}-%
\frac{5}{8}\mu ^{2}\right]
\end{equation}

There is a general relativistic contribution of $\simeq \alpha /R$.

\qquad Substituting the metric elements in the expression for $\tau _{e}$
and carrying out the calculations to the same order (with $l\simeq R\Phi $)

\begin{eqnarray}
\tau _{e} &\simeq &2\frac{r_{1}\Phi }{c}\sqrt{\frac{\left( \mu \kappa
\right) ^{2}+\left( 1-\mu \right) }{1-\mu +2\mu \kappa \beta -\beta ^{2}}} \\
\tau _{e} &\simeq &2\frac{l}{c}\left[ 1+\frac{1}{2}\beta ^{2}\right]
\end{eqnarray}%
Note that there is no first order general relativistic contribution.
However, in the difference in the time of flight between the two arms

\begin{equation}
\delta \tau _{re}=\tau _{r}-\tau _{e}\simeq \frac{l}{c}\left( \mu -2\beta
^{2}-\frac{5}{4}\mu ^{2}\right)
\end{equation}%
there is a general relativistic contribution.

\subsection{Light beam traveling along the $\protect\theta -$direction}

The light ray travels from the beam splitter along the local meridian,
either towards the North or South. In this case $dr\simeq 0$. The travel
time from the beam splitter to the end mirror is

\begin{equation}
t=-\dint\limits_{0}^{\Phi }\frac{g_{t\phi }}{g_{tt}}\frac{d\phi }{d\chi }%
d\chi \pm \dint\limits_{0}^{\Phi }\frac{1}{g_{tt}}\sqrt{g_{t\phi }^{2}\left( 
\frac{d\phi }{d\chi }\right) ^{2}-g_{tt}\left( g_{\theta \theta }+g_{\phi
\phi }\left( \frac{d\phi }{d\chi }\right) ^{2}\right) }d\chi
\end{equation}%
where the angle $\chi $ is complementary to $\theta $. The integral limits
are chosen as $0$ and $\Phi $ instead of $\chi _{1}$and $\chi _{2}$ because $%
\Phi $ is the angle subtended by the interferometer arms. The $g_{\mu \nu }$%
's maintain the values they have at the equator because we approximate the
rays as straight lines on a sphere of radius $R$. The rays have a space
trajectory

\begin{equation}
\phi =k\chi
\end{equation}%
The time of flight is

\begin{eqnarray}
t_{N} &=&-\frac{g_{t\phi }}{g_{tt}}k\Phi +\frac{1}{g_{tt}}\sqrt{g_{t\phi
}^{2}k^{2}-g_{tt}\left( g_{\theta \theta }+g_{\phi \phi }k^{2}\right) }\Phi
\\
t_{N} &\simeq &\frac{1}{g_{tt}}\sqrt{-g_{tt}g_{\theta \theta }}\Phi -\frac{%
g_{t\phi }}{g_{tt}}k\Phi
\end{eqnarray}%
Also

\begin{eqnarray}
\Omega t_{N} &=&k\Phi \\
k &=&\frac{\Omega }{\Phi }t_{N}
\end{eqnarray}%
Thus%
\begin{equation}
t_{N}=\frac{\sqrt{-g_{tt}g_{\theta \theta }}}{g_{tt}+g_{t\phi }\Omega }\Phi
\end{equation}%
The reverse path, calculated with a ($-)$ sign outside the radical in
(1.51), is just the same i.e., $t_{S}=t_{N}$, thus the total time of flight
is%
\begin{eqnarray}
t_{M} &=&t_{S}+t_{N}=2\frac{\sqrt{-g_{tt}g_{\theta \theta }}}{%
g_{tt}+g_{t\phi }\Omega }\Phi \\
t_{M} &\simeq &2\frac{l}{c}\left[ 1+\frac{1}{2}\mu +\frac{3}{8}\mu ^{2}%
\right]
\end{eqnarray}%
In terms of proper time

\begin{eqnarray}
\tau _{m} &=&t_{M}\sqrt{1-\mu +2\mu \kappa \beta -\beta ^{2}} \\
\tau _{m} &=&2\frac{l}{c}\left[ 1-\frac{1}{2}\beta ^{2}\right]
\end{eqnarray}%
Notice again the absence of general relativistic contribution; this holds
true for light travel time on the tangent or $\theta -\phi $ plane.

The differences in proper times are 
\begin{eqnarray}
\delta \tau _{rm} &=&\tau _{r}-\tau _{m}=\frac{l}{c}\mu \left( 1-\frac{5}{4}%
\mu \right) \\
\delta \tau _{em} &=&\tau _{e}-\tau _{m}=2\frac{l}{c}\beta ^{2}
\end{eqnarray}%
Expressions (1.61) and (1.62) suggest the experiment.

It is worth emphasizing that the differences are in proper time; the result
is independent of any coordinate system. Indeed a careful calculation using
isotropic or harmonic Schwarzschild coordinates confirms this assertion as
does a calculation using an expansion about a point.

These results are not really new; they conform to other relativistic time
delay phenomena such as the radar-echo experiment of Shapiro as well as the
clock corrections needed for GPS navigation satellites.

\subsection{Numerical values}

For the Milky Way Galaxy (assuming it is a homogeneous disk):

\begin{eqnarray}
\frac{\alpha }{R} &\simeq &\frac{10^{16}m}{2.8\times 10^{20}m}\simeq 10^{-4}%
\text{ to }10^{-6}; \\
\beta &\simeq &10^{-3} \\
a &=&\frac{R^{2}\Omega }{2c}=R\frac{v_{p}}{2c}\sim \left( 10^{14}\text{ to }%
10^{18}\right) m \\
\frac{a}{R} &\sim &10^{-6}\text{ to }10^{-2}
\end{eqnarray}%
for peripheral velocity $v_{p}\sim 600km/s.$

For an interferometer located on Earth of length 10cm $\delta \tau
_{rm}\simeq 6\times 10^{-15}$seconds. This corresponds to an apparent
increase in length due to the Galactic center of $\simeq 1000\unit{%
\text{\AA}%
}$ in the radial arm.

Corresponding figures for the Sun-Earth system are:

\begin{eqnarray}
\frac{\alpha }{R} &=&\frac{10^{3}m}{10^{11}m}\simeq 10^{-8}; \\
\beta &\simeq &10^{-4}
\end{eqnarray}

And at Earth's surface:

\begin{eqnarray}
\frac{\alpha }{R} &=&\frac{10^{-2}m}{6\times 10^{6}m}\simeq 10^{-8}; \\
\beta &\simeq &10^{-6}
\end{eqnarray}

The Galaxy is the major source of metric perturbation in the vicinity of
Earth.

Note that the Galactic influence may need to be built into the clock rate
adjustment in the GPS navigation system; currently only the much smaller
Earth's effect is built-in.

Why is there a measurable effect proportional to the potential $\alpha /r$,
when one can always define a new set of coordinates to make $\alpha /r=0$ at 
$r$? The answer lies in the extended reach of the interferometer; it is not
possible, except in flat space, to make the potential zero everywhere over
an extended region. The interferometer measures a potential \textit{average}
over a $10cm\times 10cm\times 10cm$ space-like region. This is not zero; and
cannot be made zero everywhere within the region using a single
transformation.

An alternative explanation is this: Two null vectors are transported
simultaneously, in closed paths, along two orthogonal axes; when they return
to the origin there is an angle defect between them, which is, as expected,
a first order effect in $\alpha /r$.

\subsection{Experiment, \ Apparatus and Noise}

As the interferometer rotates with Earth the output measures alternately, $%
\delta \tau _{rm}$ and $\delta \tau _{em}$. The signal appears as
alternating bright and dark regions in the combined beam. The interferometer
is located at the equator (assumed only to simplify the discussion), with
the two arms aligned West-East and North-South. As the interferometer
rotates with Earth, the North-South arm maintains its relative alignment
(this is not strictly true but we make this assumption to simplify the
discussion) with the Galactic center while the orientation of West-East arm
alternates between the radial and azimuthal directions every 12 hours. The
signal is sinusoidal, modulated with a period of 12 hours; it can be
recovered using a homodyne detector. This is a null experiment in the sense
that the appropriate phase of the dynamic output is zero in the absence of
any general relativistic effect, the output measures only deviations from a
flat metric.

It is worth emphasizing that the computed general relativistic fringe shift
is for \textit{each round-trip} of the light beam. The fringe shift is the
result of transporting, in a closed loop, a null vector in curved space.
Each subsequent round-trip adds to the fringe shift from the previous cycle.
By contrast a fringe shift in flat space is not cumulative; it is static.

Each round-trip, which takes $0.67\times 10^{-9}$seconds, contributes $1/5$
of a fringe shift. Thus the fringe shift is wiped out every 5 round-trips or
in $3\times 10^{-9}$seconds. In order to observe the fringe shift one needs
to strobe the output signal with a frequency which is the inverse of $%
0.67\times 10^{-9}$seconds or $1.5$ GHz.

Anisotropic Lorentz contraction of the interferometer arms (1.47) will occur
with a magnitude $\propto \left( 1/2\right) \beta ^{2}\approx 10^{-7}$. This
is smaller than the expected signal but also $90^{\circ }$ out of phase with
it, hence distinguishable.

Although we have assumed that the length of both interferometer arms is
exactly equal, this will not be so in practice, nor is it necessary. Any
remnant inequality will show up as a zero-offset in the sinusoidal signal.

A time delay of $10^{-16}$ secs. corresponds to $\simeq 1/5$ of a fringe
shift. Is this measurable in the presence of noise from sources such as
mechanical/thermal noise, photodetector noise; mechanical distortions such
as sag and tilt, seismic noise, laser intensity fluctuations, laser
wavelength/frequency fluctuations etc.\cite{Blair}? \ The effect we are
looking for is independent of wavelength, so neither laser wavelength nor
intensity fluctuations nor drift will affect the result if the apparatus is
kept in a vacuum environment. However, the interference condition does
depend on wavelength, so the laser wavelength needs to be stable to $\ll 1/5$
of a wavelength.

Among the noise sources listed, the most relevant is thermal stability since
we are looking for a change in apparent length. Low-expansion substances
such as Zerodur or sapphire would be suitable platform materials. With
expansion coefficients of $\ \simeq 10^{-6}$ a length of 10cm can be kept to
within $\ll \pm 1000\unit{%
\text{\AA}%
}$ by controlling the temperature to within $\pm 10^{-3}$K. This is a modest
challenge at room temperatures.

Vibration isolation is effective with increasing resonance frequencies; this
also makes sapphire or Zerodur suitable materials because of their
mechanical stiffness.

Integration intervals over several days or weeks will require reference
oscillators stable over long periods. Oscillators are available with
fractional frequency stabilities of $\pm 10^{-16}$ over months.

One may ask why this relatively large effect was not detected in experiments
going back to that of Michelson and Morley. We have studied published
accounts \cite{will} and have concluded, tentatively, that the apparatuses
and methods used were neither sensitive to nor designed to measure what we
are proposing.

\subsection{Proposed experiments}

Two experiments are proposed, and a recommendation:

1) Measure mass of Galaxy ( the part within the Galactic center and Solar
system)

2) Possible improvement in the uncertainties in the Shapiro time delay using
the effect of the Sun on the interferometer.

3) An immediate recommendation: Make clock rate corrections to satellite
based navigation systems.

The second would be a challenge; it may require a cryogenic environment
where temperature stabilities of $\pm 10^{-6}$K are possible.\ This,
combined with an expansion coefficient of sapphire at 4K which is estimated
to be $\sim 10^{-11}K^{-1}$, may reduce thermal stability to $\sim 10^{-17}$%
\cite{Blair}.

The Galactic center would provide a strong background for the second
experiment. It may be possible to extract the solar signal by monitoring the
interferometer output during months of June and December when the Sun, Earth
and Galactic center are aligned, then during March and September when they
form a right angle triangle. The two signals, with different amplitudes,
will have a relative phase which will progress from $0^{\circ }$ to 90$%
^{\circ }$ between December and March.

It would appear that the Galaxy can be weighed with a table-top device which
is sensitive to metric perturbations over cosmic distances.

\end{document}